# Event Loops as First-Class Values

## A Case Study in Pedagogic Language Design

Joe Gibbs Politz[a], Benjamin S. Lerner[b], Sorawee Porncharoenwase[c], and Shriram Krishnamurthi[c]

a   University of California, San Diego
b   Northeastern University
c   Brown University

**Abstract**   The World model is an existing functional input-output mechanism for event-driven programming. It is used in numerous popular textbooks and curricular settings.

The World model conflates two different tasks–the definition of an event processor and its execution–into one. This conflation imposes a significant (even unacceptable) burden on student users in several educational settings where we have tried to use it, e.g., for teaching physics.

While it was tempting to pile on features to address these issues, we instead used the Scheme language design dictum of removing weaknesses that made them seem necessary. By separating the two tasks above, we arrived at a slightly different primitive, the *reactor*, as our basis. This only defines the event processor, and a variety of execution operators dictate how it runs.

The new design enables programmatic control over event-driven programs. This simplifies reflecting on program behavior, and eliminates many unnecessary curricular dependencies imposed by the old design. This work has been implemented in the Pyret programming language. The separation of concerns has enabled new curricula, such as the Bootstrap:Physics curriculum, to take flight. Thousands of students use this new mechanism every year. We believe that reducing impedance mismatches improves their educational experience.

## The Art, Science, and Engineering of Programming

## 1 Reactive Programs in Education

Reactive programming [3] is useful for building a variety of interactive systems, including games and simulations.[1] These are widely used in education, both to motivate students (e.g., games as a way of inducing student interest in computation) and to teach specific subject matter (e.g., to simulate a predator-prey population, a physics experiment, an economics model, a PID controller, etc.). Whole languages and development frameworks, such as NetLogo [28], have been built to facilitate writing such programs.

Our focus in this paper is on a specific framework called "World" [12], which we explain more in section 2. The World model is used in several educational settings, including textbooks such as *How to Design Programs* [13], *Picturing Programs* [5], and *Programming and Programming Languages* [19], and curricula such as those of the Bootstrap outreach program [6, 7, 8]. Thus, it is used with students typically ranging in age from 12 to 25. The model was developed in Racket and has since been ported to other languages, including OCaml and Pyret.

In the World model, the user-supplied parts of programs are intended (or, with language support, forced) to be purely functional. This achieves high fidelity with the (pure) functions used to model phenomena in other disciplines, and is even used to teach the very concept of functions. In particular, when the educational goal is to teach mathematical functions, it is important that the programming functions behave similar to those of math, to avoid confusing students. Thus, while some issues discussed below can be solved differently using side-effects in client programs, such solutions would be inappropriate and even antithetical in several educational settings that use this model.

In what follows, we will refer to the programmers as "students" and their programs as "student programs". This does not preclude non-students from using World, and indeed many (including the authors) do. Nevertheless, we use these terms to remind readers that the intended users are usually beginners learning to program, so solutions that are readily evident to experienced programmers may not be apparent, meaningful, or even decipherable to beginners.

## 2 The World Programming Model

A typical reactive program matches figure 1. The external world generates a series of events: clock ticks, keystrokes, mouse movements, network packet arrivals, etc. To respond to the events of interest, the program registers a series of callbacks or *handlers*.[2] Each callback runs for a while, then returns control, possibly with some response, at which point the program is ready for the next stimulus. (For simplicity, we do not consider concurrent processing (section 9).)

---

[1] In the rest of this paper, we will use the term *simulation* to include one that has interaction, and can thus be presented as a simple game.
[2] In this paper we do not go into other semantic models such as functional reactivity. We discuss this in section 8.

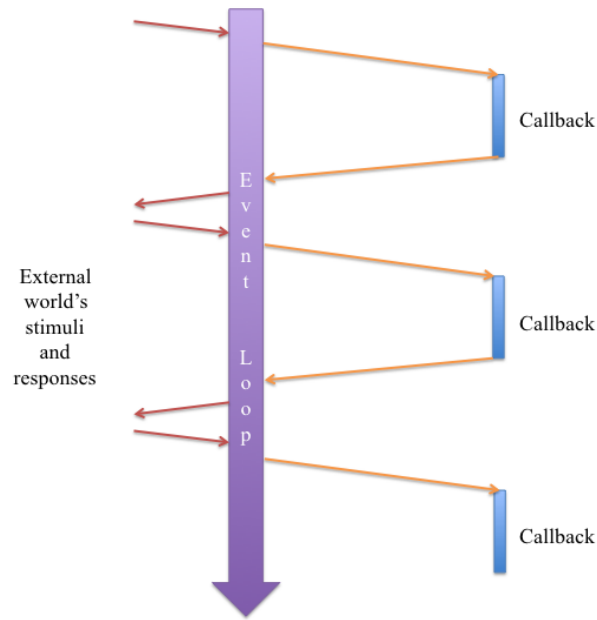

**Figure 1** A model of reactive programming.

Because the execution of "the program" has been fragmented, programmers must confront how to pass information or persist state between the fragments. In many conventional libraries, this is done using side-effects. In the World model, every program chooses a representation for each *world* (there may be several over the course of a run), which is (at least) the information that needs to be stored between handler invocations. One can think of it as the model in the sense of Model-View-Controller [16], or as the information recorded in a checkpoint for recovery.[3]

Most handlers consume the world as a parameter and return an updated world as their result. That is, they represent the computational essence of the problem. The tasks of responding to external events; scheduling them; calling the handlers; storing the updated world and passing it on to the next handler; and so on–tasks that are either administrative in nature or require knowledge of specialized and perhaps complex APIs–are built into the World framework, so the student programmer does not need to know how to do them. In particular, in contrast to many other introductory approaches, the student does not write an "outer event loop", so they write quite sophisticated programs without having to understand looping constructs (which induce many misconceptions in students [20]).

---

[3] In a way, the term "world" is misleading, because it needs to reflect only essential information, not the *entire* state; anything that is redundant or static can be left out. However, the term makes more sense in its original context [12], where it was a precursor to the "Universe" model, where worlds communicate to implement distributed programs.

**Event Loops as First-Class Values**

■ **Listing 1**   A World program that maintains a counter and draws circles of that size.

```
 1  fun time-handler(w :: Number) -> Number:
 2    w + 1
 3  end
 4
 5  fun key-handler(w :: Number, k :: String) -> Number:
 6    ask:
 7      | k == "i" then: w + 10
 8      | k == "m" then: w - 10
 9      | otherwise: w
10    end
11  end
12
13  fun drawer(w :: Number) -> Image:
14    circle(w, "solid", "blue")
15  end
16
17  fun stopper(w :: Number) -> Boolean:
18    w > 100
19  end
20
21  big-bang:
22    init: 0,
23    on-tick: time-handler,
24    on-key: key-handler,
25    to-draw: drawer,
26    stop-when: stopper
27  end
```

Listing 1 shows a simple World program (runnable at https://tinyurl.com/reactor-paper-counter-big-bang).[4,5] All the programs are written in Pyret,[6] an educational language inspired by Racket, OCaml, and Python. In this program, the initial value of the world is `0`. On every clock tick (by default, 28/second) this value increments by one. If the user presses either the `i` or `m` keys, the value of the world goes up or down by `10`, respectively.[7] Every time the world's value changes, the event loop invokes the **to-draw** handler (if provided), obtains an image, and refreshes the screen to show this image. The program stops when the **stop-when** handler (if provided) returns `true`. At that point, the entire expression evaluates to the value of the last world. The types[8] of these operations are illustrative. Given a type W to stand for the program's world representation, here are some of them:

---

[4] All tinyurl.com URLs were last accessed 2019-02-01.
[5] The actual syntax of **big-bang** is different (appendix A). We render it more readably here.
[6] Appendix B describes how readers can run every program in this paper.
[7] If the **otherwise** (or "else") clause were not included and the user were to hit any other key, the program would fall through the conditions and, in Pyret, halt with an error. Thus, the **otherwise:** w pattern in this setting means "ignore all other keys".
[8] Pyret supports both static type-checking and run-time checking of assertions as contracts.

| | | | |
|---|---|---|---|
| the value given to | `init` | has type | `W` |
| the value given to | `on-tick` | has type | `W -> W` |
| the value given to | `on-key` | has type | `W, Key -> W` |
| the value given to | `stop-when` | has type | `W -> Boolean` |
| the value given to | `to-draw` | has type | `W -> Image` |

The example above shows how a program can respond to stimuli. The programming model naturally extends to other stimuli as well, and on some platforms [29] supports sensations like GPS and tilt (e.g., `on-tilt`, with type `W, Tilt -> W`). Similarly, it can generate other forms of output, such as producing SMS messages (`to-sms`, with type `W -> String`). In both the input and output case, the student program does not itself "read" or "write"; instead, all interaction is virtualized and instead performed by the World event loop.

This design has several beneficial consequences:

1. Because of this virtualization, all the functions written above are pure, and can be tested without resort to setup, teardown, etc., i.e., no differently from any mathematical function. This feature is not only exploited by but is central to the pedagogy of all the curricula listed in section 1.

2. As programs (such as games) become more complex, they naturally require more information to be stored in the world. This motivates the introduction and learning of increasingly sophisticated datatypes and data structures. For instance, a game that goes from having one danger (e.g., an enemy or hazard) to an unbounded number is a natural way to motivate the need for sets or lists.

3. Different semantically equivalent worlds can be used to motivate computational complexity. For instance, suppose a program's goal is to compute a list of points (which is contained in its world) and to present these visually. A natural `to-draw` would iterate over all these points and generate the visual afresh each time. If, however, the number of points keeps growing by a constant on each iteration, this representation results in quadratic behavior in the number of points. An alternative is to keep the image itself in the world–even though it is redundant and can be reconstructed–and only add the new points on each iteration, thereby trading space for time to turn the quadratic into a linear-time solution (or constant-time per iteration).

## 3 Weaknesses of the World Model

Unfortunately, the simplicity of the World model is also a weakness: it is too simple. It is not surprising that, in the enormous space of all possible behaviors, there would be many behaviors that are difficult to write in this model. What is much more problematic is that there are many *natural* programs that are difficult to write. Worse, these directly interfere with and force a re-ordering of the curriculum, at which point the model becomes a hindrance rather than an aid. We present some examples below. Section 5, specifically section 5.1 and section 5.4, present solutions to these problems.

**Event Loops as First-Class Values**

- Students want to run a simulation for a limited number of ticks. For instance, they may make a prediction that a system will be in a certain state after *n* ticks, and want to test it.
  Because **big-bang** starts execution and keeps running until an external interrupt or a **stop-when** handler forces it to stop, controlling the number of steps must be done by the student program. That means students must extend the state to also track the number of steps so far, update this in every event processor, and add or alter the **stop-when** handler to check the number of steps.
  Curricularly, this adds two constraints. First, the students may not even have been introduced to **stop-when**, but are now forced to confront it. Second, students may not yet know how to create a data structure (such as a pair) to hold multiple pieces of information: if their program state was only one atomic datum, now they must learn basic data structures just to add execution control to their program.
  Finally, this solution is conceptually ugly. In most cases, information like the number of steps is outside the *domain model*; it's simply a debugging or testing aid–i.e., it's metadata about a particular *run*, not about the *problem*. Thus, from the perspective of understanding the essence of a domain, it is extraneous, and forces a poor modeling practice.
- Students in a physics class write simulations of objects in a two-dimensional space to study the laws of motion. A natural state representation is a pair of numbers, representing the object's location. At the end of the simulation, the students would like to graph the object's trajectory.
  The graphing task requires that the program keep track not only of the object's current position but the entire history of its past positions. Because this is information that needs to persist, the student needs to learn how to represent such a data structure. They will not only have to wrangle with nested structures (a pair, one element of which is *another* pair representing the object's location, the other element being the history), they will have to learn how to represent the history. This requires learning an entirely new, unbounded data structure (such as a list). This is not only a topic that was unnecessary until this point, it may not even be part of the curriculum (e.g., many curricula teach numerous physical simulations without ever covering lists). Furthermore, the student must learn the programming pattern of accumulating history, and use it correctly (updating the current location while appending the *previous* location to the history).
- Several curricula using the World model emphasize the writing of test cases, especially as examples to understand the problem statement before working on the solution code. The World model lends itself very well to testing, because all the handlers are pure functions and the program's state is virtualized. Thus, each function can be tested individually, as can their composition: e.g., in https://tinyurl.com/reactor-paper-counter-big-bang,

```
1 time-handler(10) is 11
2 key-handler(10, "i") is 20
3 time-handler(time-handler(time-handler(0))) is 3
4 key-handler(time-handler(key-handler(0, "i")), "m") is 1
```

Note, however, what is and isn't being tested. We are testing the *functions*, but we are not testing a *simulation*. If we knew that the simulation was (a) guaranteed to terminate and (b) did not require any interaction, we could run it to completion and test its final value without any manual intervention; even then, we would not be able to test intermediate states. Absent these two properties, the student would have to manually add execution control (as discussed above), and if the simulation involves interaction, we would have to find a way to supply the stimuli.

It is worth remembering that the audience for this work is novice programmers, some as young as middle-school. Thus, the workarounds suggested above are neither obvious nor easy to implement to novices; in some cases they form a significant component of the rest of the course; in others they are even outside its scope (especially so if the course is not a computing course but one in another discipline like physics, so that teaching data structures is not a topic of interest at all).

## 4 Reactors: A Small Change for a World of Difference

There are several possible analyses of the weaknesses described above. For instance, perhaps `big-bang` is simply not enough on its own. Maybe we need more constructs like `big-bang`, with more parameters, and more clauses, to handle each of the situations above. We can certainly imagine various library extensions that implements each of these variations.

One complicating factor to defining these abstractions is syntax. For pedagogic reasons, Racket's `big-bang` provides a rich syntactic language for writing its clauses. This enables both more readable programs (than, say, positional arguments) and context-sensitive error-checking with student-friendly error messages (e.g., making sure no clauses were duplicated). Therefore, in a language lacking user-defined syntactic abstraction (such as Pyret), these extensions cannot be provided by a library; they must be built into the language.

However, there is perhaps a more important design principle to bear in mind. We are inspired by the Scheme design dictum [1],

> Programming languages should be designed not by piling feature on top of feature, but by removing the weaknesses and restrictions that make additional features appear necessary.

Our proposed proliferation of operators and clauses certainly seems to inspire a search for a better design.

The key lies in our previous analysis (section 3): many of the extensions are not properties of the model per se but are related to *operations we want to perform to runs of a simulation*. That suggests separating the definition of models from the manipulation and processing of runs.

Why are we not able to easily perform operations to runs? It's because we have no *programmatic* control over a run: the `big-bang` operator both

- *defines* a reactive computation by binding handlers, and
- immediately *runs* it.

**Event Loops as First-Class Values**

This is analogous to the `let` construct found in many languages, which both defines a collection of bindings and immediately runs an associated body. But we know that `let` can be thought of as syntactic sugar for two separate operations: **lambda** and application. Can we apply the same insight here?

**Introducing Reactors**  We therefore introduce a new construct, **reactor**. The syntax of a reactor is the same as that of a **big-bang** except for the change in the keyword. However, they are quite different in behavior. A reactor simply creates a value (in Pyret, represented as an object of type `Reactor`) that embodies the simulation, without actually executing it. `Reactor` is a type constructor of one argument, which is the type of the state of the reactor. Most importantly, just as a **lambda** closes over its environment, a reactor closes over its simulation state. Two reactors can have the exact same handlers but be closed over different state values, and hence behave differently.

Thus, the **big-bang** expression of listing 1 would instead be turned into a definition (https://tinyurl.com/reactor-paper-counter-reactor),

```
r = reactor:
  init: 0,
  on-tick: time-handler,
  on-key: key-handler,
  to-draw: drawer,
  stop-when: stopper
end
```

which binds `r` to a value of type `Reactor<Number>`. A reactor can be run as below.

**Running Reactors**  The most natural thing to do with a reactor is to run it:[9]

```
interact :: Reactor<A> -> Reactor<A>
```

A user can extract the state of the reactor with

```
get-value :: Reactor<A> -> A
```

Why does `interact` return a `Reactor<A>` rather than an `A`? It's because other reactor-executing operations (shown below) have this type, so this ensures a consistent interface. The result reactor is closed over its state, so that one can resume it in that state and continue to interact with it. Consequently, in `interact(interact(r))` (where `r` is bound to some reactor), the inner call starts running `r` and keeps doing so until either its **stop-when** makes it halt or the user interrupts it. Either way, it returns a reactor closed over the last state before termination. The outer invocation resumes this reactor in that state. That means, if the previous reactor had halted not by user interaction but because of the **stop-when** clause, then–because the computations are expected to be pure–the second invocation will immediately halt, returning a reactor in the same state.

---

[9] Every operation below is available both as a function and as a method on reactors. We will use both forms in this paper.

**Tracing Reactors**   We can also *trace* a reactor, i.e., get information about its execution. Pyret defines several operations for this:

```
interact-trace :: Reactor<A> -> Table
simulate-trace :: Reactor<A>, Number -> Table
```

`interact-trace` is like `interact`, but its output is not the final state value (encapsulated in a reactor). Rather, it is a table of two columns, `tick` and `state`. This provides the state at each tick. `simulate-trace` is like `interact-trace` but with two major differences: it eliminates the visual display (i.e., runs "headless")—and therefore cannot receive any user-generated events either–and allows the computation to be limited to a certain number of ticks.

```
start-trace :: Reactor<A> -> Reactor<A>
stop-trace :: Reactor<A> -> Reactor<A>

get-trace :: Reactor<A> -> List<A>
get-trace-as-table :: Reactor<A> -> Table
```

   `start-trace` transforms the given reactor to be a "tracing" reactor, so that it logs all subsequent events until a `stop-trace`. `get-trace` retrieves the list of logged events, while `get-trace-as-table` retrieves them as a table. Therefore, `interact-trace` is equivalent to

```
get-trace-as-table(interact(start-trace(r)))
```

**Stepping Reactors**   The reason `interact` returns a reactor, not just a naked value, is for uniformity and compositionality: many other operations also consume and return reactors. In particular, every "single-stepping" operation does this, returning a reactor that represents the same behavior but encapsulating its next state.

   The key stepping operator, `react`, accomplishes two things. First, it represents the next step of the reactor. Second, it consumes as input a *virtualized* representation of an event:

```
react :: Reactor<A>, Event -> Reactor<A>
```

where the language provides a rich language of events representing the different stimuli. Thus, a user can programmatically simulate user behavior and write tests against reactor responses.

   Recall that all the reactor operations above are functional. Thus, given the following program (we drop the type annotations–which are optional in Pyret–for simplicity), found at https://tinyurl.com/reactor-paper-incr-tests:

```
fun increment(x): x + 1 end

reg = reactor:
  init: 0,
  on-tick: increment
end
```

the following tests pass:

**Event Loops as First-Class Values**

```
1  check:
2    get-value(reg) is 0
3    r2 = react(reg, time-tick)
4    get-value(r2) is 1
5    get-value(reg) is-not 1    # is-not is also a testing primitive
6    get-value(reg) is 0        # still 0 because reactors are immutable
7  end
```

**An Abstraction of Event Loops**   In short, reactors are a linguistic embodiment of an event loop. They represent the responses intended for different stimuli, bundled into a first-class value. This value can be applied to real events generated by the external world, or it can be applied to virtualized events generated manually or even programmatically. This gives the programmer the full ability to control and inspect the system's behavior while also obtaining working programs that perform real input-output. By providing straightforward support for operations like stepping, it both simplifies and encourages not only programming but testing reactive systems.

## 5   Illustrative Examples

The previous section outlines reactors and their library. Here we discuss several uses of reactors that have been enabled by the above changes, most of which are actively employed in various curricula.

### 5.1  Testing Sequences of Interactions

We return to the example program in listing 1. As noted earlier (section 3), in the World model, it is difficult to test for the behavior of the system as a whole. In contrast, with a reactor, it is easy to test *runs* of a reactor. Indeed, we can write a useful helper abstraction in the language itself (https://tinyurl.com/reactor-paper-counter-reactor):

```
1  fun after-n-ticks<T>(rctr :: Reactor<T>, n :: Number) -> Reactor<T>:
2    if n <= 0:
3      rctr
4    else:
5      after-n-ticks(rctr.react(time-tick), n - 1)
6    end
7  end
```

which simulates running a reactor for n ticks. We can then write the following (successful) tests for our example program:

```
1  check:
2    after-n-ticks(r, 10).get-value() is 10
3    r.react(time-tick).react(time-tick).react(keypress("i")).get-value()
         ↪ is 12
4    r.react(keypress("i")).react(keypress("m")).react(time-tick).get-value
         ↪ () is 1
5  end
```

which is exactly what we rued not being able to test easily.

The last of these tests is particularly notable. If the only way to test a reactor was manually, we may never have thought to test for behavior where a keypress precedes the very first tick. And even if it we had thought of it, it would take extreme dexterity to be able to perform two keypresses before even the first tick has elapsed (or, in general, between any two ticks $\frac{1}{28}$ of a second apart). Virtualizing the passage of time eliminates these problems.

### 5.2 Pausing, Interacting, and Debugging at the REPL

We don't only have to use `react` in tests: it is also useful at the REPL.

Suppose we have built a simulation and find something odd in its behavior. When we see the odd behavior occur, we can stop the program either using the language's Stop button or clicking the "x" to close the reactor's output window or some other such mechanism. This produces a value: the last (state of the) reactor.

If we bind this reactor to a name (say `game-r`), we can use the testing methods to move the game along and examine the output. This is a technique we commonly use in teaching. For instance, we use the following sequence of interactions in the REPL:

```
> r2 = game-r.react(time-tick)
> game-r.draw()
> r2.draw()
```

The first line advances the game by a tick and names this reactor. The second line invokes the **to-draw** handler of `game-r` to render an image of that game state. The third line similarly renders an image of the game state one tick later. It can be very useful to put these images side-by-side for debugging and program comprehension.

Observe that this interaction crucially depends on reactors closing over their state, and the updates being functional. Otherwise, `game-r.draw()` is likely to not produce the desired output after `.react(time-tick)` has been called. Observe also that this drives home how ticks are truly virtual, not connected to real computer time. This point will be further reinforced in section 7.

### 5.3 Comparing Alternate Models

In our physics curriculum [7], students are taught about two different ways of modeling dynamical phenomena. One is the traditional high school textbook way, where system states are expressed as functions of time. For instance, the displacement of an object under uniform linear motion might be represented as (https://tinyurl.com/reactor-paper-alternate-models):

```
init-x = 0
delta-t = 1
v = 10

fun x-at-t(t):
  init-x + (v * t)
end
```

**Event Loops as First-Class Values**

In that curriculum, these are called *(time) parametric* representations. A simulation of this can of course be written as a reactor.[10]

These are contrasted against *differential* representations, where states are made explicit and students write a function showing how states update from one tick to the next. These are very naturally represented as reactors:

```
fun next-x(x):
  x + (v * delta-t)
end

diff-r = reactor:
  init: init-x,
  on-tick: next-x
end
```

In essence, the reactor represents an integrator, adding the difference on each time interval to the previous state to obtain a new state. Note that it is natural for such a reactor to not have a **stop-when**, so testing only for its "final" value is not feasible.

Now we want students to see that these two representations model the *same resulting phenomena*.[11] We can do this naturally through tests: using after-n-ticks (section 5.1), we should be able to check for the equivalence of states in the two representations. Sure enough, the following tests pass:

```
check:
  fun test-after(steps):
    param-x = x-at-t(steps)
    diff-x = after-n-ticks(diff-r, steps).get-value()
    param-x == diff-x   # == checks for equality
  end
  test-after(10) is true
  test-after(100) is true
end
```

Indeed, we can even write[12]

```
  for each(i from range(0, 100)):
    i satisfies test-after
  end
```

to gain confidence in the lock-step equivalence of the two methods. If this were to fail, Pyret would immediately present a counter-example. Given that it succeeds, the student can then be taught the use of analytic methods to prove their equivalence.

---

[10] Though it is somewhat awkward to do so: we leave this as an exercise to the reader.

[11] By illustrating this similarity, we want them to also understand why the latter phenomenon produces the same outcome as the former, and hence motivate the differential and integral calculus.

[12] In Pyret, the common test pattern `f(x) is true` can be written more idiomatically as `x satisfies f`. Though semantically equivalent, the latter provides better diagnostic output. In case `f(x)` is in fact `false`, the former provides the unhelpful error report that `true` is not `false`. In contrast, the latter expressly reports the specific x that does not satisfy f.

### 5.4 Tracing and Replaying

Generating a trace is useful because, combined with the functional nature of the code, we can replay the steps of the reactor. Consider the following intentionally highly simplistic reactor, shown in https://tinyurl.com/reactor-paper-trace-replay:

```
fun time-handler(w): w + 10 end

fun stopper(w): w > 100 end

fun drawer1(sz):
  circle(sz, "outline", "blue")
end

r = reactor:
  init: 0,
  on-tick: time-handler,
  to-draw: drawer1,
  stop-when: stopper
 end
```

Obviously it isn't the case here, but suppose the circle were hard to find in the resulting image. This can happen due to occlusion, colors melding with the background, small objects being lost against a noisy background, etc.

It is easy to record a trace of the reactor's execution:

```
t = interact-trace(r)
```

This generates a table with the columns named `"tick"` and `"state"` holding their corresponding values at each tick.

Suppose we define a different rendering function that makes the desired portion of the image more salient, addressing whatever was creating a problem (e.g., rendering a solid version of the object and/or giving it a higher *z*-index):

```
fun drawer2(sz):
  circle(sz, "solid", "red")
end
```

Then, we can easily use the table builder functionality in Pyret to provide a table of the image views:

```
t.build-column("screenshot", lam(row): drawer2(row["state"]) end)
```

This functionally extends the table bound to t to add one new column, `"screenshot"`. Each row's screenshot is defined by applying the new rendering function (drawer2) to that row, which in turn accesses the row's state value (row["state"]) as the basis for its new rendering. A prefix of the table is below:

**Event Loops as First-Class Values**

## 5.5 Comparing Initial Conditions

Because reactors are just expressions, their creation can easily be parameterized using just ordinary functions. For instance, consider this free-fall simulation parameterized over its initial state (https://tinyurl.com/reactor-paper-init-cond):

```
1  accel = -0.3
2
3  fun ticker(w):
4    { y: w.y + w.vy,
5      vy: w.vy + accel }
6  end
7
8  fun make-sim(start-y):
9    reactor:
10     init: { y: start-y, vy: 0 },
11     stop-when: lam(state): state.y < 0 end,
12     on-tick: ticker
13   end
14 end
```

The { ... } notation is Pyret's syntax for an inline object, so the body of ticker is an expression that returns an updated state object.

We can now make two reactors parameterized over different starting heights:

```
1  sim200 = make-sim(200)
2  sim400 = make-sim(400)
```

We can then test these extensively:

```
1  check:
2    sim200.react(time-tick).get-value().vy
3      is sim400.react(time-tick).get-value().vy
4    sim200.react(time-tick).get-value().y
5      is sim400.react(time-tick).get-value().y - 200
6    after2-200 = sim200.react(time-tick).react(time-tick)
7    after2-400 = sim400.react(time-tick).react(time-tick)
8    after2-200.get-value().vy is after2-400.get-value().vy
9    after2-200.get-value().y is after2-400.get-value().y - 200
10 end
```

The tests demonstrate that, starting from zero velocity, reactors with different initial conditions have the same velocity after two steps, and their positions differ only by the initial conditions. (Though **big-bang** expressions can also be parameterized in this way, the World model lacks the execution control used in the tests above.) Observe that all of this is verifiable without having to visually run and inspect the simulations themselves.

## 6 Implementation Details

Because reactors are currently only implemented for Pyret, and Pyret is built atop JavaScript, this section is necessarily JavaScript-centric. Building atop a different platform might entail a very different implementation.

When an interaction begins, the reactor runtime installs JavaScript callbacks corresponding to the various on- handlers, each of which updates the single shared piece of state for the reactor. These calls are interleaved with calls of the to- handlers (that cause side-effects like updating the screen) after state updates, per the semantics in the World paper [12].

Once running, a reactor must both respond to events and produce a final value (if it terminates). To respond to any events in a browser tab, which is single-threaded, the reactor must relinquish control to JavaScript for events to actually be processed. Therefore, the continuation at the start of the interaction must be recorded so that, on termination, it can receive the final state. Thus, implementing reactors in JavaScript requires reifying continuations.

There is a further complication. While JavaScript programmers are sensitive to the need to write only short-running computations in callbacks (otherwise the browser window appears to hang, and the browser offers to kill the computation), student programmers neither know this nor should be exposed to this detail.[13] Therefore, Pyret uses a continuation-based abstraction to save and restore computation both *programmatically*, around interactions, and *periodically*, during individual event processing. This work has been abstracted more broadly into the Stopify system [4].

## 7 Nested Reactors

One issue we have not discussed until now is what happens if a reactor is run "inside" another reactor. What should the semantics be?

Consider the following (intentionally contrived) program. We first define the following abstraction (https://tinyurl.com/reactor-paper-nested-reactors):

---

[13] Even if they were told about it, it would be non-trivial for them to determine how to refactor their computation into lots of little pieces, and teaching this would overwhelm whatever other curricular point was being made–such as modeling physics.

```
1  fun get-digit(t :: String) -> Number:
2    r = reactor:
3      init: none,
4      to-draw: lam(_): text(t, 30, "black") end,
5      on-key: lam(_, k): string-to-number(k) end,
6      stop-when: is-some,
7      close-when-stop: true # closes window on stopping
8    end
9    r.interact().get-value().value
10 end
```

This function pops up a window with a prompt (t); the initial value doesn't matter because it will soon be replaced. The function string-to-number returns an option type (none if the string doesn't correspond to a number, and some number otherwise; observe that the reactor loops until an actual digit is entered). The .value at the end extracts the value from the some, returning a pure number.

This forms a truly encapsulated abstraction. For instance, a programmer can write:

```
1  get-digit("first digit") + get-digit("second digit")
```

This expression will pop up one window; when a digit is pressed, close this window and pop up a second one; and when another digit is pressed, evaluate to their sum. Observe that the continuation to each call to get-digit is non-trivial: for instance, that for the first call is

```
1  lam(hole):
2    hole + get-digit("second digit")
3  end
```

which demonstrates why the reactor must store a proper continuation to resume computation once the reactor completes.

Now suppose we want to drive this function from an outer reactive loop. This will repeatedly invoke get-digit, with its state recording the running sum. If the user ever enters the digit 0 this will halt the outer reactor, returning the total sum. Thus, the reactor's state is a pair: a flag indicating whether 0 has been pressed, and the running sum.

```
1  r2 = reactor:
2    init: {sum: 0, done: false},
3    on-tick:
4      lam(s-d): # stands for "sum" + "done"
5        k = get-digit("next number")
6        if k == 0:
7          {sum: s-d.sum, done: true}
8        else:
9          {sum: s-d.sum + k, done: false}
10       end
11     end,
12   stop-when: lam(s-d): s-d.done end
13 end
```

Now consider what happens when we run the outer reactor (`interact(r2)`). After it invokes the inner reactor (in `get-digit`), what happens to the outer one? The two obvious choices are:[14]

- It is still "running", which means it continues to register ticks, which means the calls to `get-digit` continue to pile up (and can overwhelm the system).
- It "suspends", and only resumes when the inner reactor terminates.

The above program is written with the latter, *modal* behavior in mind.

It is not clear to us that there is a clearly "correct" choice here. DrRacket's World system implements the former semantics. Under this, the program is clearly flawed: essentially, multiple reactors are battling for control. In this semantics, it is not easy to implement modal behavior (e.g., to pop up a modal window in a game) without the outer reactor knowing about the inner one and voluntarily "suspending" itself. Even this suspension is complicated: it would have to be in the form of a busy wait, and would require every handler to contain logic that checks whether the reactor is active and operate only if it is. Furthermore, an inner reactor cannot enforce this behavior; it would be at the mercy of the outer reactor to have been programmed with this logic.

A Racket port of the World model to the Web [29] therefore expressly chose to break with this behavior and instead cause automatic suspension, which is the same behavior that Pyret's reactors now use. We feel this causes less surprise in many settings. (After all, many user interface elements–such as file pickers–behave in this modal manner.) However, it naturally comes at a cost. A tick is now not only a virtualized but also a localized unit of time: between each pair of ticks, arbitrary many real ticks may have elapsed. Because any expression may launch a reactor, it is only safe to assume that any unknown expression did. On the other hand, there are settings (e.g., a game with a "do you want to continue?" modal popup) where this kind of suspension of virtual time is exactly what a programmer would want.

**Implementation Consequences of Nested Reactors**

Because of the chosen semantics of nested reactors, the actual implementation of reactors is a stack; each level of the stack is a pair consisting of a queue of events, and a continuation. The continuation is because we don't assume (nested) reactors are invoked in tail position, so this allows resumption of the reactor's invoking context. Furthermore, the browser's handlers are treated like registers and saved and restored on every reactor entry and exit, with care taken to ensure that (a) no handlers are present when no reactor is active, and (b) no handlers from outer reactors are present in inner reactors (e.g., in the example above, the reactor inside `get-digit` should not dynamically "inherit" an **on-tick** handler from the calling reactor, bound to `r2`).

This model obeys an important invariant: every new event appends to the event queue of the *top-most* frame. Since earlier events in the queue can trigger nested reactor invocations, however, the queue of events must be preserved, to process once the nested invocation completes. These event queues are reminiscent of, and were inspired by, the eventspaces of Racket [15].

---

[14] We ignore other choices, such as disabling some handlers but not others, since these would easily lead to violations of reactor invariants, making local reasoning about them impossible.

It is worth noting that this model is quite different from the flat, continuation-less JavaScript/DOM model. Therefore, it does not map directly to JavaScript but must instead be implemented explicitly. On a platform–such as Racket–with eventspaces, continuations, and so on, the implementation would be more direct and far simpler.

## 8  Related Work

Our work is not a traditional research contribution: we do not claim any significant technical novelty. Most of the technical work that our design builds on is already cited in relevant parts of the paper.

Naturally, the World style is not new; variations of it have been considered throughout the history of integrating imperative computation in functional programming. A history of Haskell [18, section 7.1] says that its designers considered a model

> in which the state of the world is passed around and updated, much as one would pass around and update any other data structure in a pure functional language. This "world-passing" model was never a serious contender for Haskell, however, because we saw no easy way to ensure "single-threaded" access to the world state.

We avoid this difficulty almost entirely because the event loop ensures this single-threading (at the cost of some flexibility). Of course, it is conceivable that an individual function could "stash away" its world value, but this would require imperative state, which we eschew by enforcement or fiat. A more sophisticated solution is of course to have the language check this property, which is the approach taken in Clean [2]. Another approach is to make copies of worlds' state and carefully attend to the logic around copying changes from one to another, called "sprouting" and "committing" in the World model of Warth and Kay [27].

Elm's notion of `Program` makes the idea of world-passing style central to its model-view-update architecture [11, The Elm Architecture].[15] A `Program` is built from an initial state, a functional updater over messages and state, and a function that produces a view from a state. `Program`s can be created "headless" for testing purposes by using a different constructor.[16] An instance of `Program` represents an entire program that is instantiated and managed by Elm's runtime system, typically through a JavaScript function call (outside of Elm). The extra-lingual startup and the difference between the "headless" and "real" program is an interesting contrast with reactors, where the same value in-language allows for both the "headless" and interactive operation, and all the usual in-language facilities can be used to operate on traces and resulting reactor programmatically. A library with these features could likely be built atop Elm's toolkit and enjoy many of the benefits of reactors.

---

[15] https://guide.elm-lang.org/architecture/, last accessed 2019-02-01.
[16] https://package.elm-lang.org/packages/elm-lang/core/latest/Platform#Program, last accessed 2019-02-01.

There are of course other approaches to programming for reactive systems [3], including doing so functionally (e.g., functional reactivity). However, functional reactivity can be a poor fit in our setting. Briefly, behaviors are extremely elegant but require students to understand a more complex operational semantics. Events require some facility with a widespread use of functions as values, when many of our students are still struggling to learn first-order functions. (The use of special syntax for worlds and reactors masks the first-class use of functions.) Furthermore, functional reactivity is designed to enable reactive programs to be written in a more deeply nested and compositional manner, but some of our curricula are trying to *introduce* nesting and composition, and hence cannot assume it as a skill students have and want to embrace. The models described in this paper have instead functioned well in our settings, though undoubtedly there are other educational contexts that would prefer functional reactivity.

Several other recent systems use the term "reactor" to refer to first-class values that interact with events. The `Reactor[T]` of Prokopec and Odersky [22] is intended to serve as part of an extension of the actor programming model, orchestrating communication between multiple concurrent reactors, which is not the focus of this paper. With Pyret's reactors extended to handle concurrency–most likely in the style of Universe programs (section 9)—a direct comparison between these approaches would be more meaningful. Similarly, the reactors of Van den Vonder, Myter, De Koster, and De Meuter [26] are designed to link the parallelism of WebWorkers with reactive programming frameworks in the browser, an implementation technique that may be useful for concurrent extensions of the reactors in this paper. The `define-reactor` and `rho` forms of the Haii language [21] are designed for constructing dataflow graphs. The programming model of Haii eschews traditional functions in favor of reactors "all the way down," while the reactors of this paper are explicitly designed to work with traditional functions as handlers.

In *Java: An Eventful Approach*, Bruce, Dalnyluk, and Murtagh [10] also use a pedagogic programming style that involves event handlers as methods paired with drawing operations. There are two main contrasts with our work: (a) their handlers use side-effects to manage state rather than acting as functional updaters, and (b) outputs like drawing are not managed by a single handler, but can be (and are, in the book's examples) manipulated using state across multiple handlers. Many of the testing-focused examples in section 4 rely on the functional style of Pyret's reactors' handlers.

## 9 Discussion

We have discussed the design, use, and implementation of reactors, a variant of the World model of computation. Both are in widespread use in education across a wide range of ages. Reactors directly address some of the problems we have run into when using the World model. Furthermore, reactors represent a *principled separation of concerns*, and nicely reflect the Scheme design dictum (section 4).

**Event Loops as First-Class Values**

**Concurrency**   The original work on World introduced not only reactive but also distributed programming [12] through a Universe model. A Universe is essentially a set of Worlds running on (potentially) different run-time systems. We see no difficulty in replacing the Worlds with reactors.

A less evident extension, which we have not examined, is to have reactors running in parallel. On a Web page, for instance, one could imagine several different elements each having a reactor attached and communicating with one another through message-passing. While we do not see a conceptual difficulty with this, we are unconvinced that this would be particularly elegant or illustrative. For such a program, it may make more sense to switch to a different model, such as functional reactivity.

**Levels**   A few programming systems have explored the notion of "language levels", i.e., languages that grow in complexity with a student's knowledge [9, 14, 17]. It would be interesting to consider similar growth for reactors. The simplicity of the current reactor model has worked well in many of our pedagogic settings, and it would be potentially confusing to add state, more sophisticated handlers, or more reflective features to reactors. A language level mechanism, however, would provide both a vehicle for exposing more such features, and semantic criteria that help determine which levels should be exposed when.

**Naming**   One distinct advantage of **big-bang** is that the common use-case—of defining *and immediately running* a reactor–does not require any additional ceremony. This is definitely lost with the reactor setup, where a student must write

```
1  r = reactor: ... end
2  interact(r)
```

In contrast to **big-bang**, this
- forces the student to pick a name for the reactor,
- forces them to come up with a second name if they want to define more than one,[17]
- indents the code of the reactor a little more,
- demands that they remember the name of one more library function, and
- introduces several new possible places where they might make an error.

These may all seem like minor or non-issues to trained programmers, but they are genuine points of trouble or at least work for beginners. Worse, it forces them to spend some of their precious cognitive load budget on inessential matters.

One solution is to write an expression of the form

```
1  interact(reactor: ... end)
```

---

[17] Indeed, we have found that in our own programming practice–just as in this paper!—we routinely call the first reactor r instead of choosing some semantically meaningful name. The second reactor is often r2, and only as the number grows do we consider renaming them to something meaningful.

but for most beginners (and even many advanced programmers), this is likely not easy to read and follow (since ... is likely several lines long). Observe that this is the moral equivalent of writing

```
1 ((lambda (...) ...) ...)
```

Another solution is, of course, to provide **big-bang** as a macro over **reactor** that hygienically creates and immediately uses a fresh name. Naturally, because that name would not later be visible, the other needs (section 3) that motivate reactors (such as tracing) become difficult or impossible to perform.

A somewhat different solution entirely to the general problem of syntax overhead is taken by the Bootstrap:Algebra curriculum [6], wherein students improve their mathematics ability [23, 24] through writing a videogame. This material works under a very tight time constraint (since it is injecting programming into a high-stakes subject) and is often taught to quite young students. In this setting, students do not write a **big-bang** (or **reactor**) *at all*.

Instead, they write functions that follow a particular naming convention: for instance, the function that updates the position of the player in the game must be called update-player, and likewise for other game elements.[18] The curriculum support library then automatically harvests these named functions and constructs a **big-bang** expression with the necessary handlers connected appropriately to the student's functions. While this approach of course only works in limited settings, it does circumvent the naming issues entirely for beginners, and offers a way to ameliorate the concerns described above. Indeed, the follow-up Bootstrap:Reactive curriculum [8] then explicitly pulls back the curtain and shows students how to write the full program for themselves.

**Semantics Visualization**   Reactors (like World) have a potentially complex semantics. In particular, the trade-off for writing pure functions is that they impose what we call an *agency* problem: instead of actually reading a key, the student provides a function that receives the read key; instead of actually drawing on the screen, the student provides a function that generates a drawing that will be shown on the screen. It is an open problem to conduct user studies that learn just how much of a problem this is in practice.

One potentially helpful tool would be a presentation–say visual–of the program's behavior, which (roughly speaking) in educational contexts is called a "notional machine" [9]. A presentation just the level of the language would not suffice: it would likely show a collection of disembodied function calls without clear connection between them. Rather, we need a notional machine that clearly shows the behavior of the reactor abstraction itself, and these calls within the context of that abstraction. We hypothesize that any cognitive burdens created by the agency problem are likely to be significantly ameliorated through a user-centric design of a notional machine for reactors.

---

[18] This idea is reminiscent of the approach taken by Java Beans and other nominal reflection-based libraries.

**Implications for Lambda**   This work was inspired by an analogy: the relationship between `let` and lambda. On reflection, we find that reactors are even more like a lambda than we perhaps initially imagined. They are values that encapsulate code; they are by default inert; when triggered, they execute that code; they are instantiable; and each instance has its own local state. Thus, reactors are very much like a lambda.

However, reactors also do much more than a lambda: they can be single-stepped, they can be traced to reveal their state, and so on, and all these are *linguistic* primitives. In most languages, lambdas do not offer these features: the features can *at best* be obtained using debuggers or other external tools, and the results are not under programmatic control (Siskind and Pearlmutter's `map-closure` [25] is a notable exception). Of course, in some settings revealing the state of a lambda would lead to dangerous security violations, so we do no claim this should be done indiscriminately; furthermore, the notion of a "step" is well-defined in the reactor semantics [12] but not so clear for lambdas. Nevertheless, we do wonder about the design of an alternate lambda mechanism that has more reactor-like behaviors.

**Acknowledgements**   We thank Danny Yoo, whose work on Whalesong inspired some of our decisions and whose implementation work we used as a starting point. We are deeply grateful to our physics curriculum collaborators, especially Rebecca Vieyra and Colleen Megowan-Romanowicz, for working with us on our experiments with reactors. We are grateful to our reviewers for a close reading, thoughtful comments, and helpful related work pointers. This work was funded by multiple US National Science Foundation grants.

## A  `big-bang` Syntax

In Racket, **big-bang** is a new language construct ("special form") with custom syntax. For instance, the example in listing 1 would be written as

```
(big-bang 0
  (on-tick time-handler)
  (on-key key-handler)
  (to-draw drawer)
  (stop-when stopper))
```

Because Pyret does not have syntactic abstraction support, **big-bang** was written using a *function* interface. Therefore, the relevant part of listing 1 would actually be written:

```
big-bang(0,
  [list:
    on-tick(time-handler),
    on-key(key-handler),
    to-draw(drawer),
    stop-when(stopper)])
```

This is arguably "simpler" because it's made up of existing language primitives and does not require learning any new syntax at all. However, the result is visually much more complex; it forces students to confront lists before they may be ready for it and certainly before they may need it; and it forces all well-formedness checks (such as the kinds of handlers) to be performed at run-time. The reactor syntax appears to us more visually pleasant, avoids the unnecessary data structure dependency, and checks the handler constructs for errors statically. (Of course, it means the list of handlers cannot be assembled dynamically, but we have never found a use for that power.)

## B  How to Run Programs

All the programs in this paper can be run directly as written in Pyret. Pyret can be run from the desktop (https://github.com/brownplt/pyret-lang/) or directly in the browser from https://code.pyret.org/. Since different code fragments require different library support, it is best to include the following two lines at the top of every file:

```
include reactors
include image
```

The one exception is listing 1, which requires the syntax translation described in appendix A, and the header lines

```
include world
include image
```

## About the authors

**Joe Gibbs Politz** (jpolitz@eng.ucsd.edu) is an Assistant Teaching Professor in Computer Science and Engineering at the University of California, San Diego.

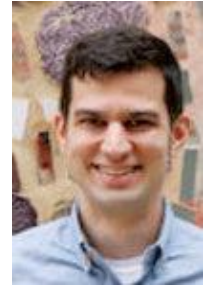

**Benjamin S. Lerner** (blerner@ccs.neu.edu) is an Assistant Teaching Professor in the College of Computer and Information Sciences at Northeastern University.

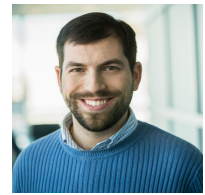

**Sorawee Porncharoenwase** (sorawee.pwase@gmail.com) is now a PhD student at the University of Washington.

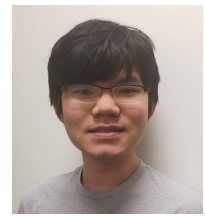

**Shriram Krishnamurthi** (sk@cs.brown.edu) is the Vice President of Programming Languages (no, not really) at Brown University.

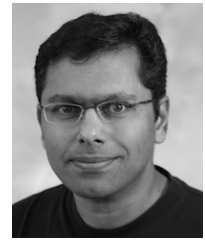